\documentclass[aps, prb, reprint, superscriptaddress]{revtex4-1}

\usepackage{todonotes}
\usepackage{bm}
\usepackage{graphicx}
\usepackage{amsmath}
\usepackage{mathbbol}
\usepackage{amssymb}
\usepackage[normalem]{ulem}
\usepackage{color}
\usepackage{todonotes}

\graphicspath{{./figures/}}

\begin{document}

\title{Stable and manipulable Bloch point}

\author{Marijan Beg}
\email{marijan.beg@xfel.eu}
\affiliation{European XFEL GmbH, Holzkoppel 4, 22869 Schenefeld, Germany}
\affiliation{Faculty of Engineering and Physical Sciences, University of Southampton, Southampton SO17 1BJ, United Kingdom}
\author{Ryan A. Pepper}
\author{David Cort\'{e}s-Ortu\~{n}o}
\author{Bilal Atie}
\author{Marc-Antonio Bisotti}
\author{Gary Downing}
\affiliation{Faculty of Engineering and Physical Sciences, University of Southampton, Southampton SO17 1BJ, United Kingdom}
\author{Thomas Kluyver}
\affiliation{European XFEL GmbH, Holzkoppel 4, 22869 Schenefeld, Germany}
\author{Ondrej Hovorka}
\affiliation{Faculty of Engineering and Physical Sciences, University of Southampton, Southampton SO17 1BJ, United Kingdom}
\author{Hans Fangohr}
\email{hans.fangohr@xfel.eu}
\affiliation{European XFEL GmbH, Holzkoppel 4, 22869 Schenefeld, Germany}
\affiliation{Faculty of Engineering and Physical Sciences, University of Southampton, Southampton SO17 1BJ, United Kingdom}

\begin{abstract}
The prediction of magnetic skyrmions being used to change the way we store and process data has led to materials with Dzyaloshinskii-Moriya interaction coming into the focus of intensive research. So far, studies have looked mostly at magnetic systems composed of materials with single chirality. In a search for potential future spintronic devices, combination of materials with different chirality into a single system may represent an important new avenue for research. Using finite element micromagnetic simulations, we study an FeGe disk with two layers of different chirality. We show that for particular thicknesses of layers, a stable Bloch point emerges at the interface between two layers. In addition, we demonstrate that the system undergoes hysteretic behaviour and that two different types of Bloch point exist. These `head-to-head' and `tail-to-tail' Bloch point configurations can, with the application of an external magnetic field, be switched between. Finally, by investigating the time evolution of the magnetisation field, we reveal the creation mechanism of the Bloch point. Our results introduce a stable and manipulable Bloch point to the collection of particle-like state candidates for the development of future spintronic devices.
\end{abstract}

\maketitle
In recent years, materials with the Dzyaloshinskii-Moriya interaction (DMI)~\cite{Dzyaloshinsky1958, Moriya1960} came back into the focus of intensive research, mostly due to the discovery of magnetic skyrmions~\cite{Rossler2006, Muhlbauer2009} that promise to revolutionise the way we store and process data~\cite{Fert2013}. Contrary to the ferromagnetic exchange which tends to align all spins parallel to each other, the DMI energy is minimised when neighbouring spins in the crystal lattice are perpendicular. Their mutual competition results in a twist between neighbouring spins which eventually leads to the helical magnetic order~\cite{Bak1980} over the entire crystal lattice. Although the angle between neighbouring spins, and consequently the helical period, depend on the relative strength of DM compared to the ferromagnetic exchange energy, the handedness of the emerged helical order in a helimagnetic material depends on the material's chirality - sign of the DM energy constant. So far, studies have looked mostly at magnetic systems composed of materials with single chirality. In a search for potential future spintronic devices and new particle-like magnetisation states, combination of materials with different chirality may represent an important new avenue for research.

\begin{figure}
  \includegraphics{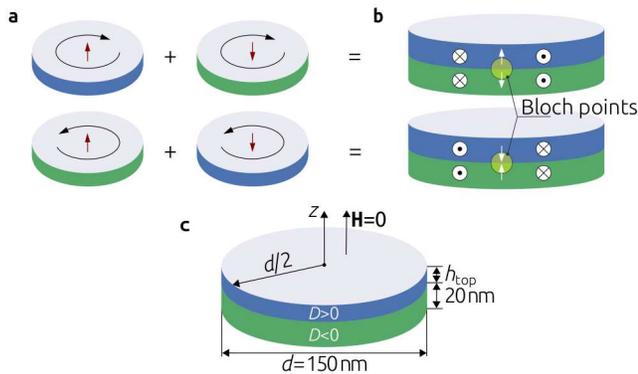}
  \caption{\label{fig:geometry} \textbf{A thought experiment of stacking magnetic vortices to obtain a Bloch point.} (\textbf{a})~Magnetic vortex can have two different chiralities and two different polarisations, which results in four possible states for a magnetic vortex. (\textbf{b})~By stacking two disks hosting magnetic vortices with the same handedness and different polarisation results in a magnetisation configurations containing a discontinuity, also known as a Bloch point. (\textbf{c})~Geometry of the system we are going to study in this work.}
\end{figure}

To guide our investigation we start with a thought experiment. In the absence of chiral interactions, in thin film disks, a vortex configuration can emerge due to the competition between ferromagnetic exchange and demagnetisation energies. A vortex can have two different kinds of handedness as well as two different core polarisations. This results in four possible states of a magnetic vortex in a disk as shown in Fig.~\ref{fig:geometry}~(a). In order to avoid possible cacophony with the terminology established in previous studies, throughout this work, we refer to the $+z$ or $-z$ core orientation of the vortex state as its polarisation, and to the clockwise or counterclockwise chirality/winding as its handedness. Let us now imagine we have two disks where vortices with the same handedness were formed, but with different polarisation. We stack these on top of each other as though they were parts of a much thicker disk as shown in Fig.~\ref{fig:geometry}~(b). This would result in a magnetisation configuration that is continuous at all points of the sample, except at the centre, where two vortex cores point in the opposite direction~\cite{Miltat2011}. This discontinuity in the magnetisation field is called a Bloch point~\cite{Feldtkeller1965, Doring1968}. There are two possible discontinuous polarisations of the cores, and consequently two different types of a Bloch point, as we show in Fig.~\ref{fig:geometry}~(b). A Bloch point has a high magnetic energy, especially because of the high ferromagnetic exchange energy at the point where two spins point in the opposite direction. Consequently, the entire system tends to reduce its energy by expelling the high energy Bloch point out of the sample. In our system, this would occur by changing the polarisation of either of the two vortices so that both vortex cores have the same polarisation but keep the same handedness.

In the system of two stacked disks we analysed, the reorientation (change of polarisation) of one of the vortices to expel the Bloch point is possible because in a magnetic vortex both polarisations are allowed for a particular handedness. However, if the Dzyaloshinskii-Moriya energy is present in the system, the polarisation and handedness of a vortex (more precisely, vortex-like state) are always related to each other and only two possible states of a magnetic vortex are allowed~\cite{Beg2015}. In our thought experiment, this would imply that changing the vortex core polarisation while keeping the handedness fixed in order to expel the Bloch point from the system is not possible. Therefore, we can assume that a Bloch point will remain present in the sample. In this work, we explore this idea.

Using finite element micromagnetic simulations, we demonstrate the existence and stability of a Bloch point at the interface of two materials with different chirality. In addition, by exploring the hysteretic behaviour, we find two different Bloch point configurations and show how by using an external magnetic field, one can switch between them. Finally, by analysing the time evolution of a uniformly magnetised system, we reveal the creation mechanism of a Bloch point. We believe that the existence and stability of a Bloch point in the studied system, as well as the possibility of its manipulation, introduces a new particle-like state to the collection of candidates for the development of future spintronic devices.

\section{Results}

\subsection{Stability}

\begin{figure*}
  \includegraphics{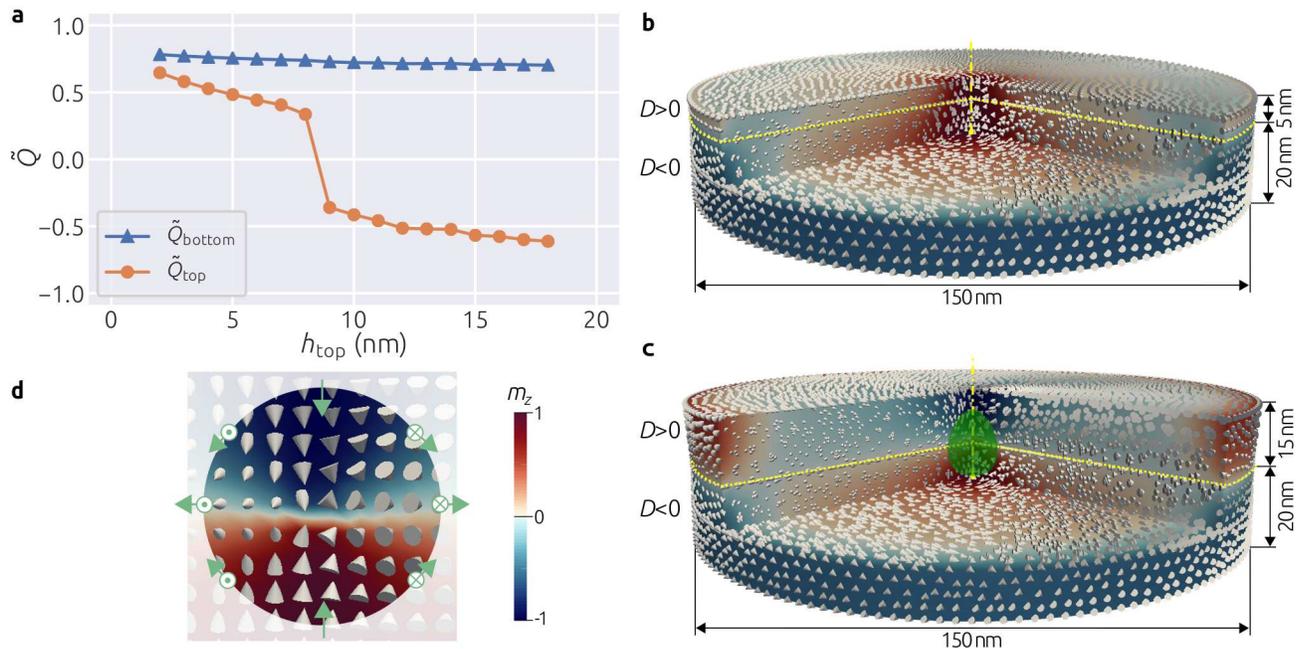}
  \caption{\label{fig:stability} \textbf{Normalised topological charge $\tilde{Q}$ dependence in individual layers, identified equilibrium states, and a Bloch point structure.} (\textbf{a})~Dependence of the normalised topological charge $\tilde{Q}$ in individual layers on $h_\text{top}$. Two identified magnetisation configurations: (\textbf{b})~for $h_\text{top} \le 8 \,\text{nm}$ and (\textbf{c})~for $h_\text{top} \ge 9 \,\text{nm}$. (\textbf{d})~Bloch point structure and demonstration that the magnetisation direction covers a sufficiently small closed surface surrounding the Bloch point exactly once.}
\end{figure*}

We begin our study by exploring whether a Bloch point emerges in the system shown in Fig.~\ref{fig:geometry}~(c), obtained as a result of a thought experiment we conducted in the introduction. In order to base our study on realistic material parameters and therefore encourage the experimental verification of our predictions, we use cubic B20 FeGe material parameters (see Methods section). We set the DMI constant in the bottom layer to be negative ($D=-|D|$) and fix its thickness $h_\text{bottom}$ to be $20 \,\text{nm}$. In the top layer we impose a positive DMI constant ($D=|D|$) and vary its thickness $h_\text{top}$ between $2 \,\text{nm}$ and $18 \,\text{nm}$ in steps of $1 \,\text{nm}$. The diameter of the disk sample is $150 \,\text{nm}$ and no external magnetic field is applied. For each thickness of the top layer, we begin by initialising the system with a uniform magnetisation configuration in the positive out-of-plane ($+z$) direction. This initial configuration corresponds to the fully saturated magnetisation configuration feasible in an experimental setup. After that, we relax the system by integrating a set of dissipative time-dependent equations until the condition of vanishing torque ($\mathbf{m} \times \mathbf{H}_\text{eff}$) is satisfied. After the relaxed (equilibrium) state is obtained, we compute a three-dimensional topological charge in the top and the bottom layer individually using
\begin{equation}
  \label{eq:topological_charge_3d}
  Q = \frac{1}{4\pi} \int \mathbf{m} \cdot \left( \frac{\partial \mathbf{m}}{\partial x} \times \frac{\partial \mathbf{m}}{\partial y}\right) \,\text{d}^{3}r,
\end{equation}
as suggested by Lee et al.~\cite{Lee2009}. Because $Q$ is proportional to the volume (and accordingly the thickness) of individual layers, we normalise it by the thickness $\tilde{Q} = Q/h$, in order to make the values mutually comparable between different layers. The only reason we use normalised topological charge is to track changes in the magnetisation of individual layers in the system. We show in Fig.~\ref{fig:stability}~(a) how normalised topological charges $\tilde{Q}$ of the equilibrium state in individual layers depend on the thickness of the top layer $h_\text{top}$. Although the topological charge of the bottom layer $\tilde{Q}_\text{bottom}$ remains relatively constant, a sharp decrease of $\tilde{Q}_\text{top}$ between $8 \,\text{nm}$ and $9 \,\text{nm}$ thicknesses of the top layer is evident. Consequently, we assume that depending on the thickness of the top layer, two different configurations can emerge as equilibrium states in the studied system. Subsequently, we explore these two magnetisation configurations.

To explore the state that emerges for $h_\text{top} \le 8 \,\text{nm}$, we visualise the magnetisation obtained for $h_\text{top} = 5 \,\text{nm}$ in Fig.~\ref{fig:stability}~(b) and for $h_\text{top} \ge 9 \,\text{nm}$ magnetisation configurations, we show in Fig.~\ref{fig:stability}~(c) the state obtained with $t_\text{top} = 15 \,\text{nm}$. We show the magnetisation vector field together with the $m_{z}$ scalar field after we cut out a slice from the sample in order to be able to look inside and inspect its centre. We observe that for both states, vortex-like configurations with clear chiralities and polarisations emerge in individual layers. In the literature, the vortex-like states formed as a consequence of DM energy are also called quasi-ferromagnetic~\cite{Rohart2013}, edged vortex~\cite{Du2013}, or incomplete skyrmion~\cite{Beg2015, Carey2016, Pepper2018} states. In both states we identified, the polarisation and the handedness of the vortex-like states in bottom layers are the same and related by the right hand rule - characteristic for the negative DM energy constant. On the other hand, the vortex-like states in the top layer differ for two equilibrium states we obtained.

The vortex-like state in the top layer with $5 \,\text{nm}$ thickness shown in Fig.~\ref{fig:stability}~(b), has the same handedness and polarisation as the thicker bottom layer. However, this is against the left hand rule which establishes the relationship between the handedness and polarisation for a positive DMI constant. Therefore, the DM energy density in the top layer is significantly higher than in the bottom layer and one can ask why the polarisation of the vortex-like state in the top layer is not reversed in order to minimise its energy. Because the energy contribution of the top layer to the total energy of the system is relatively small due to its thickness, it is energetically cheaper for the top layer to follow the magnetisation of the bottom layer (violate the left hand rule, and eventually have the higher energy density), than to reorient itself, which would result in an energetically more expensive Bloch point at the interface between two layers.

Now, we look at magnetisation configuration in the system with $h_\text{top} = 15 \,\text{nm}$, shown in Fig.~\ref{fig:stability}~(c). The chiralities of vortex-like states in both layers are the same, but their polarisations are opposite and a Bloch point emerges between them. For $h_\text{top} \ge 9 \,\text{nm}$ and above, the energy contribution of the top layer to the total energy of the system becomes significant. Therefore, the top layer now does not follow the polarisation of the bottom layer, but reverses its polarisation to follow the left hand rule. It is now energetically cheaper for the system to host a Bloch point than to tolerate the top layer following the polarisation of the bottom layer and having high energy density.

The location marked by a circle in the middle of the sample in Fig.~\ref{fig:stability}~(c) identifies a Bloch point: a noncontinuous singularity in the magnetisation vector field where the magnetisation magnitude vanishes to zero~\cite{Feldtkeller1965,  Doring1968}. Because micromagnetic models assume constant magnetisation magnitude, the precise magnetisation configuration at the Bloch point cannot be obtained using micromagnetic simulations~\cite{Andreas2014}. However, it is known how to identify the signature of a Bloch point in such situations: the magnetisation direction covers any sufficiently small closed surface surrounding the Bloch point exactly once~\cite{Slonczewski1975, Thiaville2003}. We illustrate this property in Fig.~\ref{fig:stability}~(d).

\subsection{Hysteretic behaviour}

In the previous section we demonstrated the existence and stability of a Bloch point in the proposed system at zero external magnetic field. In all relaxation simulations so far, we started the relaxation from the uniform magnetisation configuration in the positive out-of-plane ($+z$) direction which resulted in the vortex-like state in the bottom layer with the $+z$ polarisation. Furthermore, for $h_\text{top} \ge 9 \,\text{nm}$, top layer relaxed in the vortex-like configuration with negative $-z$ polarisation. This resulted in a head-to-head Bloch point as we showed in Fig.~\ref{fig:stability}~(d). At this point, we can assume that if we started our relaxations from the uniform configuration in the $-z$ direction, the polarisations of the bottom and the top layer would be in $-z$ and $+z$ directions, respectively. Consequently, a tail-to-tail Bloch point would emerge. Motivated by this assumption and knowing that uniform magnetisation configurations can be obtained by applying a sufficiently strong external magnetic field, we now explore the hysteretic behaviour of the proposed system in order to verify our assumption.

\begin{figure}
  \includegraphics{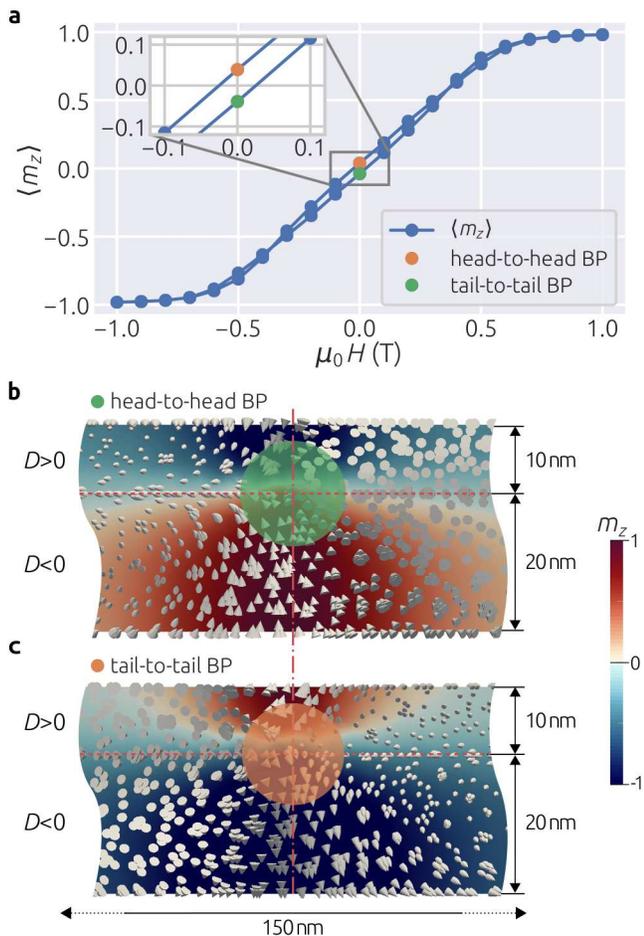}
  \caption{\label{fig:hysteresis} \textbf{Hysteretic behaviour and two different Bloch point configurations.} (\textbf{a})~Average out-of-plane magnetisation $\langle m_{z} \rangle$ as a function of external magnetic field $\mu_{0}H$. Two Bloch point configurations at zero external magnetic field: (\textbf{b})~head-to-head and (\textbf{c})~tail-to-tail.}
\end{figure}

We fix the thickness of the bottom layer with $D<0$ to be $20 \,\text{nm}$ as before and set the thickness of the top layer with $D>0$ to be $10 \,\text{nm}$ so that we can expect a Bloch point to emerge. The diameter of the disk is $150 \,\text{nm}$. We apply an external magnetic field in the out-of-plane $+z$ direction and vary it between $1 \,\text{T}$ and $-1 \,\text{T}$ in steps of $0.1 \,\text{T}$. We simulate the hysteretic behaviour in the standard way by relaxing the system to an equilibrium state after changing the external magnetic field, and then using the resulting state as the starting point for a new energy minimisation. We show the average out-of-plane magnetisation component $\langle m_{z}\rangle$ as a function of an external magnetic field $\mu_{0}H$ in Fig.~\ref{fig:hysteresis}~(a). From this plot, we see that hysteretic behaviour of the simulated system is evident and now we investigate what are the equilibrium magnetisation states at zero external magnetic field.

We start our hysteresis simulations with $1 \,\text{T}$ external magnetic field, which saturates the sample magnetisation in the $+z$ direction. By reducing the external magnetic field in steps of $0.1 \,\text{T}$, we reach $\mu_{0}H=0$ and plot the equilibrium magnetisation vector field together with $m_{z}$ scalar field in the $xz$ cross section containing the disk sample centre in Fig.~\ref{fig:hysteresis}~(b). We see that bottom and top layers relax in vortex-like magnetisation configurations with the same handedness, but with $+z$ and $-z$ polarisations, respectively. This results in a head-to-head Bloch point as we observed in the previous section. Now, we keep reducing the external magnetic field and when we reach $-1 \,\text{T}$, the sample is again fully saturated, but this time in the $-z$ direction. Starting from this magnetisation state, we increase the field in same steps until we reach $\mu_{0}H=0$, and plot the $xz$ cross section of the equilibrium state in Fig.~\ref{fig:hysteresis}~(c). Bottom and top layers are now hosting vortex-like states with the same handedness, but now with $-z$ and $+z$ polarisations, respectively. Therefore, a tail-to-tail Bloch point configuration is formed. Finally, we keep increasing an external magnetic field until we reach the $1 \,\text{T}$ value from which we started the hysteresis loop simulation. We show the evolution of the magnetisation field in the $xz$ cross section through the entire hysteresis loop in Supplementary Video 1. We demonstrated that the studied system undergoes hysteretic behaviour and that two different types of Bloch point configurations can emerge. In addition, by applying an external magnetic field, one can manipulate the Bloch point and switch between head-to-head and tail-to-tail configurations.

\subsection{Creation mechanism}

\begin{figure}
  \includegraphics{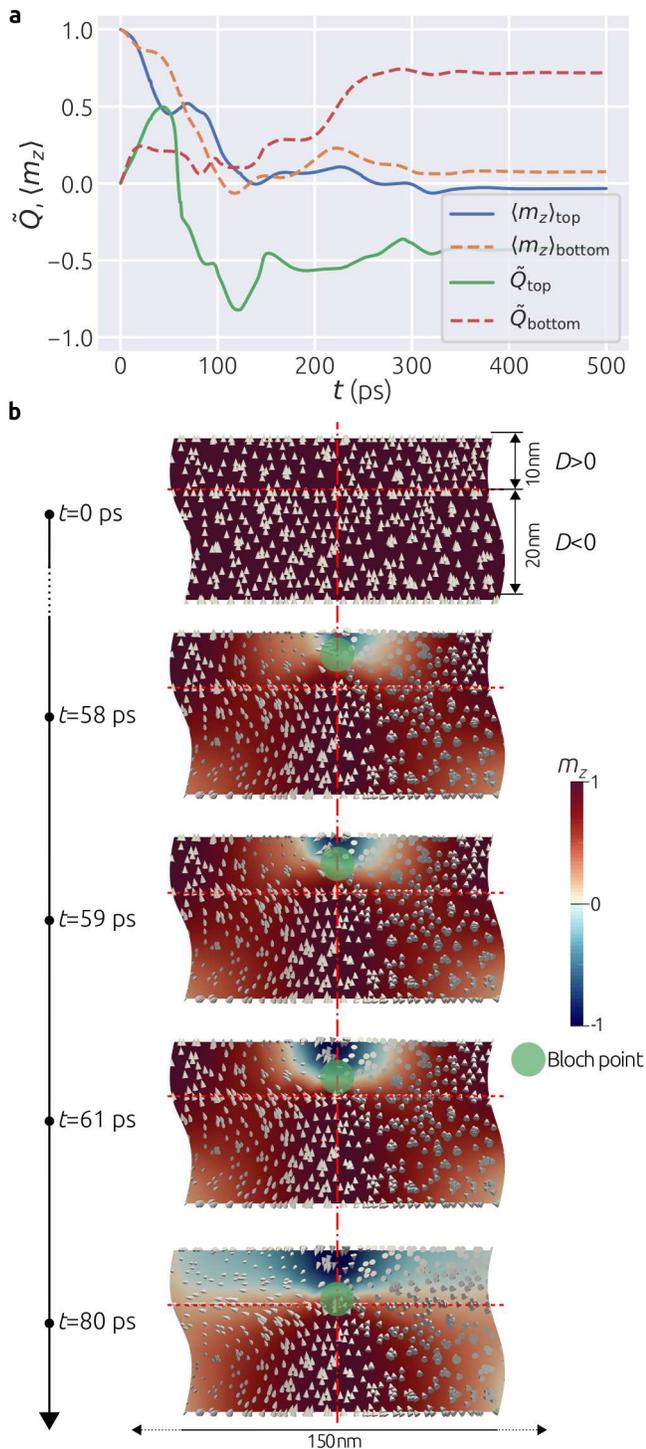}
  \caption{\label{fig:creation} \textbf{Time evolution and the Bloch point creation mechanism.} (\textbf{a})~Time evolution of the normalised topological charge $\tilde{Q}$ and the average out-of-plane magnetisation component $\langle m_{z} \rangle$ for the bottom and top layers individually. (\textbf{b})~Snapshots of the magnetisation configuration in the $xz$ cross section containing the sample centre at different points in time.}
\end{figure}

Now, we explore how the Bloch point is created in the sample. We simulate the same sample we used in hysteresis simulations, but this time we do not apply an external magnetic field. We start by initialising the system in the positive out-of-plane $+z$ direction. After that, we let the magnetisation evolve towards its equilibrium state and record the magnetisation configuration every $1 \,\text{ps}$. We set the Gilbert damping value to the experimentally measured value~\cite{Beg2017} $\alpha=0.28$. In Fig.~\ref{fig:creation}~(a) we show time evolutions of the normalised topological charge $\tilde{Q}$ as well as the average out-of-plane magnetisation component $\langle m_{z} \rangle$ for both bottom and top layers individually.

Starting from $\tilde{Q}_\text{bottom} = \tilde{Q}_\text{top} = 0$ and $\langle m_{z} \rangle_\text{bottom} = \langle m_{z} \rangle_\text{top} = 1$ at $t=0 \,\text{ps}$, which corresponds to the initial uniform state, all quantities evolve. In order to explore how the Bloch point is created in the sample, we show snapshots at different points in time in Fig.~\ref{fig:creation}~(b). We show the magnetisation vector field together with the $m_{z}$ scalar field in the $xz$ cross section containing the centre of the sample.

By locating a Bloch point in the magnetisation fields (denoted in magnetisation field plots) in Fig.~\ref{fig:creation}~(b), we explore its creation mechanism. Starting from the uniform configuration at $t=0 \,\text{ps}$, until approximately $t = 55 \,\text{ps}$ the magnetisations in both layers evolve so that vortex-like states with the same handedness and polarisation are formed. At $t=58 \,\text{ps}$ we identify that the polarisation of the top layer begins to reverse and a Bloch point is formed at the top boundary of the sample. Furthermore, Bloch point propagates downwards - towards the interface between layers, until it eventually reaches its final position at $t=80 \,\text{ps}$. We notice that the Bloch point creation mechanism is based on the polarisation (core orientation) reversal of the top layer which is the same as the reversal mechanism for skyrmionic states in confined helimagnetic nanostructures~\cite{Beg2015}. More precisely, in the process of reversal of the top layer polarisation, the Bloch point is created in the same way - at the boundary, propagating through the thickness of the sample and sitting at the interface between layers rather than being expelled out of the sample. We show the entire creation mechanism in the $xz$ cross section of the sample in Supplementary Video 2.

\section{Discussion}

Using finite element micromagnetic simulations, we studied nanostructures composed of two layers with different handedness. We found that for certain thicknesses of individual layers, a stable Bloch point can emerge at the interface between the two layers. By applying an external magnetic field we found that the system undergoes hysteretic behaviour and two different types of Bloch point configurations exist. In the first, the magnetisation at the centre of each layer points towards the interface - the head-to-head configuration. In the second, the magnetisation points away from the interface towards the outer surface of the disk - the tail-to-tail configuration. We demonstrated the switch between these two different types of Bloch point using an external magnetic field. Finally, we simulated the time evolution in order to determine the mechanism by which the Bloch point is created, showing that the Bloch point is created at the boundary of the thinner layer and then propagates through the thickness of the sample until it reaches the boundary between layers where it sits. Now, we discuss our findings focusing on four main topics: (i) the possibility of fabrication, (ii) the suitability of micromagnetic simulations for the study of Bloch points, (iii) similarities and differences of our studied system with previous works, and (iv) implications of our results.

We based our simulations on realistic FeGe material parameters~\cite{Beg2015} for a realistically sized disk in order to encourage the experimental verification of our predictions. In addition, in all of our simulations we used a uniform magnetisation configuration as an initial magnetisation state which is feasible to obtain in an experimental setup by applying a strong field. More precisely, applying an external magnetic field of approximately $1 \,\text{T}$, and then removing it, should allow the system to relax into the magnetisation configuration containing a Bloch point. However, before our predictions can be validated, one should be able to grow a helimagnetic thin film which consists of two layers with different chirality. More precisely, in the growth of a thin film, it is necessary to be able to switch the Dzyaloshinskii-Moriya energy constant $D$ once a layer of sufficient thickness has been grown. Recent research by Spencer at al.~\cite{Spencer2018} demonstrated that by changing the Co content $y$ in Fe$_{1-y}$Co$_{y}$Ge compound grown on Si(111), it is possible not only to modify the value of the DMI constant, but also to change its sign. This finding may contribute to the growth of a thin film system, composed of two layers with different chirality in order to fabricate our proposed system hosting a stable and manipulable Bloch point.

In this work, we used finite elements to discretise the continuous magnetisation fields. Because of the imposition of a constant magnetisation saturation in micromagnetics, this method is inaccurate to determine the Bloch point structure~\cite{Andreas2014}. This is because at the Bloch point the magnetisation vanishes to zero~\cite{Feldtkeller1965,  Doring1968}. However, it is well known that micromagnetics can provide us enough information to determine whether the Bloch point is present in the sample and its position, by analysing the magnetisation field on a sufficiently small closed surface surrounding the Bloch point. More precisely, it is necessary for the magnetisation direction to cover a sufficiently small closed surface surrounding the Bloch point exactly once~\cite{Slonczewski1975, Thiaville2003}. Using this condition we were able to determine the existence and the position of a Bloch point. However, because of the imposed constant magnetisation saturation, we did not attempt to determine its internal structure, which was not the topic of this work. As an example of suitability of micromagnetics to determine the existence of a Bloch point, Rybakov et al.~\cite{Rybakov2015} predicted, using micromagnetic simulations, the existence of a chiral bobber - a hybrid particle emerging at the surface of chiral magnets composed of both smooth magnetisation configuration and a Bloch point. The existence of a Bloch point as a part of chiral bobber was then experimentally confirmed in FeGe~\cite{Ahmed2018, Zheng2018}, demonstrating the validity of the micromagnetic predictions.

To some extent similar concept was introduced by Zhang et al.~\cite{Zhang2016}, where two ferromagnetic layers hosting skyrmions were antiferromagnetically coupled in order to suppress the skyrmion Hall effect, when skyrmions are driven using spin-polarised currents. However, in that system, no Bloch point can emerge between two skyrmions with different polarisations because they are physically separated by an insulating spacer of non-zero thickness. On the other hand, we achieve different polarisations (in our case, of vortex-like states) by stacking helimagnetic layers with different chirality. Because of that, there is no insulating spacer between the two layers with different chiralities and a Bloch point can be identified.

Our discovery of a stable and manipulable Bloch point in a planar magnetic system could add the Bloch point to the collection of already existing and well studied particle-like magnetic configurations that could possibly change the way we store and process data. Now, we discuss and speculate about the possibilities of how the Bloch point we explored in this work can be employed in spintronic applications. We found two different Bloch point configurations that can be switched using an external magnetic field, thus one can propose that a single information bits (0 or 1) can be encoded using the type of a Bloch point (HHBP or TTBP) -- similar to the bit-patterned media. However, from the hysteresis simulations, we saw that the average out-of-plane component of magnetisation $\langle m_{z} \rangle$ at zero field is almost zero due to the symmetry of the magnetisation field. This poses a difficulty for reading such information bits. A possible solution is to increase the thickness of the bottom layer, which does not affect the stability of the Bloch point, but increases $\langle m_{z} \rangle$ at zero field. Furthermore, our initial studies show that the Bloch point does not emerge only is disk geometries, but also in a wide variety of different geometries, such as nanostrips. In this case, the manipulation of a Bloch point is not only limited to its type, but also its position in the sample. This manipulation can be achieved using an external excitation, for instance, external magnetic fields or spin-polarised currents. However, a Bloch point coupling to such external excitations is beyond the scope of this work and should be a topic of further research.

All results obtained in this work can be reproduced from the repository in Ref.~\onlinecite{Repository}, which contains micromagnetic simulation, data analysis, and plotting scripts.

\section{Methods}

In our simulations, magnetisation dynamics is governed by the Landau-Lifshitz-Gilbert (LLG) equation~\cite{Landau1935, Gilbert2004}
\begin{equation}
  \label{eq:llg_equation}
  \frac{\partial \mathbf{m}}{\partial t} = -\gamma_{0}^{*} \mathbf{m} \times \mathbf{H}_\text{eff} + \alpha\mathbf{m} \times \frac{\partial \mathbf{m}}{\partial t},
\end{equation}
where $\gamma_{0}^{*} = \gamma_{0} (1+\alpha^{2})$, with $\gamma_{0} = 2.21 \times 10^{5} \,\text{m}\,\text{A}^{-1}\text{s}^{-1}$ and $\alpha \ge 0$ is the Gilbert damping. We compute the effective magnetic field $\mathbf{H}_\text{eff}$ using
\begin{equation}
  \mathbf{H}_\text{eff} = -\frac{1}{\mu_{0}M_\text{s}} \frac{\delta E[\mathbf{m}]}{\delta \mathbf{m}},
\end{equation}
where $E[\mathbf{m}]$ is the total energy functional, and contains several energy contributions:
\begin{equation}
  \label{eq:total_energy}
  E = \int \left[ w_\text{ex} + w_\text{dmi} + w_\text{z} + w_\text{d} \right] \,\text{d}^{3}r.
\end{equation}

The first term $w_\text{ex} = A \left[ (\nabla m_{x})^{2} + (\nabla m_{y})^{2} + (\nabla m_{z})^{2} \right]$ with material parameter $A$ in Eq.~\ref{eq:total_energy} is the energy density of the symmetric exchange contribution. The unit vector field $\mathbf{m}$, with Cartesian components $m_{x}$, $m_{y}$, and $m_{z}$, represents the normalised magnetisation field $\mathbf{M} = M_\text{s}\mathbf{m}$, where $M_\text{s}$ is the saturation magnetisation. The second term $w_\text{dmi} = D \mathbf{m} \cdot \left(\nabla \times \mathbf{m} \right)$ with material parameter $D$ is the Dzyaloshinskii-Moriya (DM) energy density. This form of DM energy density is obtained by including Lifshitz invariants for materials of crystallographic classes T and O. The coupling of magnetisation to an external magnetic field $\mathbf{H}$ is defined by the Zeeman energy density term $w_\text{z} = - \mu_{0}M_\text{s}\mathbf{H} \cdot \mathbf{m}$, with $\mu_{0}$ being the magnetic constant. The $w_\text{d}$ term is the demagnetisation (magnetostatic) energy density. Because demagnetisation energy cannot be neglected in the study of FeGe helimagnetic nanostructures~\cite{Beg2015, Vousden2016, Carey2016, Beg2017} we include its contribution in all simulations. We assume the simulated material is isotropic and neglect the magnetocrystalline anisotropy energy contribution. The specific boundary conditions~\cite{Rohart2013} have been validated by a series of standard problem simulations proposed by Cort\'{e}s-Ortu\~{n}o et al.~\cite{Cortes2018}. The FeGe material parameters we use are:~\cite{Beg2015} saturation magnetisation $M_\text{s} = 384 \,\text{kA}\,\text{m}^{-1}$, exchange energy constant $A = 8.78 \,\text{pJ}\,\text{m}^{-1}$, and Dzyaloshinskii-Moriya energy constant $|D| = 1.58 \,\text{mJ}\,\text{m}^{-2}$. The relevant length scales are the exchange length $l_\text{ex} = \sqrt{2A/\mu_{0}M_\text{s}^{2}} = 9.73 \,\text{nm}$ and helical length $L_\text{D} = 4\pi A/D = 70 \,\text{nm}$. We discretise the finite element mesh so that the maximum spacing between two neighbouring mesh nodes is below $l_\text{max} = 3 \,\text{nm}$.

We implemented this micromagnetic model in the finite element method framework and developed a micromagnetic simulation tool Finmag~\cite{Finmag}. For the low-level finite element operations, we use the FEniCS project~\cite{Logg2012} and for the adaptive step time integration we use Sundials/CVODE solver~\cite{Hindmarsh2005, Cohen1996}.

\bibliographystyle{naturemag}

\begin{acknowledgments}
 This work was financially supported by the OpenDreamKit – Horizon 2020 European Research Infrastructure project (676541), EPSRC Centre for Doctoral Training grant EP/L015382/1, and the EPSRC Programme grant on Skyrmionics (EP/N032128/1).
\end{acknowledgments}

\section{Author contributions}
M.B. and H.F conceived the study, and M.B. performed finite element micromagnetic simulations. B.A., R.A.P, and D.C.-O contributed by running additional micromagnetic simulations using different micromagnetic simulation tools to confirm and validate main results. M.-A.B., M.B. D.C.-O., R.A.P., T.K, G.D., and H.F. developed the micromagnetic finite element based simulator Finmag. M.B. with the assistance of H.F., R.A.P., and O.H. prepared the manuscript.

\section{Competing interests}
The authors declare no competing financial or non-financial interests.

\end{document}